\documentclass[pra,twocolumn,amsmath,amssymb,superscriptaddress,showpacs]{revtex4-1}

\makeatletter
\def\@bibdataout@aps{%
 \immediate\write\@bibdataout{%
  @CONTROL{%
   apsrev41Control,author="08",editor="1",pages="0",title="0",year="1"%
  }%
 }%
 \if@filesw
  \immediate\write\@auxout{\string\citation{apsrev41Control}}%
 \fi
}%
\makeatother 

\usepackage[latin9]{inputenc}
\usepackage{amsmath}
\usepackage{amssymb}
\usepackage{epsfig}

\usepackage{xcolor} 

\usepackage{ucs}
\usepackage{bbm}

\usepackage{hyperref}
\hypersetup{colorlinks=true, citecolor=blue, urlcolor=blue, linkcolor=blue}

\newcommand{\ket}[1]{|#1\rangle}
\newcommand{\ketbra}[1]{| #1\rangle \langle #1|}

\newcommand{\be}{\begin{equation}}
\newcommand{\ee}{\end{equation}}
\newcommand{\eea}{\end{eqnarray}}
\newcommand{\bea}{\begin{eqnarray}}

\newcommand{\va}[1]{\ensuremath{(\Delta#1)^2}}

\newcommand{\ex}[1]{\ensuremath{\langle{#1}\rangle}}
\newcommand{\exs}[1]{\ensuremath{\langle{#1}\rangle}}

\newcommand{\qed}{\ensuremath{\hfill \blacksquare}}

\newcommand{\kommentar}[1]{}
\newcommand{\trace}{{\rm Tr}}

\newcommand{\forget}[1]{}

\newcommand{\EQ}[1]{Eq.~\eqref{#1}}
\newcommand{\EQS}[1]{Eqs.~\eqref{#1}}
\newcommand{\EQL}[1]{Equation~\eqref{#1}}
\newcommand{\SEC}[1]{Sec.~\ref{#1}}
\newcommand{\FIG}[1]{Fig.~\ref{#1}}

\newcommand{\REF}[1]{Ref.~\cite{#1}}
\newcommand{\REFS}[1]{Refs.~\cite{#1}}

\newcommand{\APP}[1]{Appendix~\ref{#1}}

\newcommand{\AVGVAR}[1]{\overline{\rm var}(\varrho)}
\newcommand{\AVGFQ}[1]{\overline{F}_{\rm Q}[\varrho]}
\newcommand{\AVGFQPURE}[1]{\overline{F}_{\rm Q}[\ket{\Psi}]}
\newcommand{\AVGFQLOG}[1]{\overline{F}^{\rm log}_{\rm Q}[\varrho]}
\newcommand{\AVGAXY}[1]{\overline{\vert {#1}_{12}\vert^2}}
\newcommand{\AVGAXX}[1]{\overline{\vert {#1}_{11}\vert^2}}
\newcommand{\AVGAKL}[1]{\overline{\vert {#1}_{kl}\vert^2}}
\newcommand{\AVGAKK}[1]{\overline{\vert {#1}_{kk}\vert^2}}
\newcommand{\AVGVECN}[1]{\int M(d\vec n)  \exs{{#1}_{\vec n}^2}}
\newcommand{\AVGV}[1]{\overline{V}(\varrho)}
\newcommand{\AVGVSHORT}[1]{\overline{V}(\varrho)}
\newcommand{\AVGfA}{\overline{f}}

\newcommand{\comment}[1]{}

\begin{document}

\title{Lower bounds on the quantum Fisher information based on the variance and various types of entropies}

\author{G\'eza T\'oth}
\email{toth@alumni.nd.edu}
\homepage{http://www.gtoth.eu}
\affiliation{Department of Theoretical Physics, 
University of the Basque Country
UPV/EHU, P.O. Box 644, E-48080 Bilbao, Spain}
\affiliation{IKERBASQUE, Basque Foundation for Science, 
E-48013 Bilbao, Spain}
\affiliation{Wigner Research Centre for Physics, Hungarian Academy of Sciences, P.O. Box 49, H-1525 Budapest, Hungary}




\begin{abstract}
We examine important properties of the difference between the variance and the quantum Fisher information over four, i.e., $(\Delta A)^2-F_{\rm Q}[\varrho,A]/4.$ We find that it is equal to a generalized variance defined in Petz [J. Phys. A 35, 929 (2002)] and Gibilisco, Hiai, and Petz [IEEE Trans. Inf. Theory 55, 439 (2009)]. We present an upper bound on this quantity that is proportional to the linear entropy. As expected, our relation shows that for states that are close to being pure, the quantum Fisher information over four is close to the variance. We also obtain the variance and the quantum Fisher information averaged over all Hermitian operators, and examine its relation to the von Neumann entropy. 
Apart from the usual quantum Fisher information, we also consider the Kubo-Mori-Bogoliubov quantum Fisher information.
\end{abstract}

\date{\today}

\maketitle

\section{Introduction}

Quantum metrology is a subfield of metrology that takes advantage of quantum phenomena to achieve a high precision in magnetometry, frequency measurements, and several other areas of interferometry \cite{Giovannetti2004Quantum-Enhanced,Paris2009QUANTUM,Demkowicz-Dobrzanski2014Quantum,Pezze2014Quantum,Toth2014Quantum,Pezze2016Non-classical,Helstrom1976Quantum,Holevo1982Probabilistic, Petz2008Quantum}. There have been successful experiments with cold gases, trapped ions and photons to create quantum states useful for high precision metrology such as spin-squeezed states \cite{Kitagawa1993Squeezed,Wineland1994Squeezed,Gross2012Spin,Ma2011Quantum,Hald1999Spin,Echaniz2005Conditions,Sewell2012Magnetic,Riedel2010Atom-chip-based,Gross2010Nonlinear}, Greenberger-Horne-Zeilinger (GHZ) states \cite{Greenberger1990Bells,Bouwmeester1999Observation,Pan2000Experimental,Zhao2003Experimental,Lu2007Experimental,Gao2010Experimental,Leibfried2004Toward,Sackett2000Experimental,Monz201114-Qubit}, symmetric Dicke states \cite{Dicke1954Coherence,Wieczorek2009Experimental,Prevedel2009Experimental,Lucke2014Detecting,Lucke2011Twin,Hamley2012Spin-nematic}, and many-body singlet states \cite{Behbood2014Generation}. Quantum metrology played a role even in the recent experiments with the squeezed-light-enhanced gravitational  wave detector GEO 600 \cite{Demkowicz-Dobrzanski2013Fundamental}. Experiments are being carried out achieving a larger and larger precision, reaching recently a 10-times improvement compared to the shot-noise limit, i.e., the best precision achieavable by uncorrelated particle ensembles \cite{Hosten2016Measurement}.

Partly due to the experimental successes, there has been a rapid theoretical development in quantum metrology. In particular, there has been a large effort to understand better the quantum Fisher information, 
which is a central notion in quantum metrology. 
It is connected to the task of estimating the phase $\theta$  for the unitary dynamics of a linear interferometer 
\be
U=\exp(-iA\theta),
\ee 
assuming that we start from $\varrho$ as the initial state, where $A$ is a Hermitian operator.
A tight bound on the precision of the phase estimation is given by the Cram\'er-Rao bound as  
\be
\va{\theta}\ge {1}/{{F}_{\rm Q}[\varrho,A]},
\ee
where $F_{\rm Q}[\varrho,A]$ is the quantum Fisher information of the state \cite{Giovannetti2004Quantum-Enhanced,Paris2009QUANTUM,Demkowicz-Dobrzanski2014Quantum,Pezze2014Quantum,Helstrom1976Quantum,Holevo1982Probabilistic,Braunstein1994Statistical, Petz2008Quantum,Braunstein1996Generalized}.  

It has been found that the quantum Fisher information is strongly connected to quantum entanglement \cite{Pezze2009Entanglement,Hyllus2012Fisher,Toth2012Multipartite,Hyllus2010Not}, which has been used to estimate multipartite entanglement by direct measurement of the sensitivity \cite{Lucke2011Twin,Krischek2011Useful,Strobel2014Fisher}. 
It has been investigated how various sets of quantum states, such as random bosonic states and states with a positive partial transpose perform metrologically \cite{Oszmaniec2016Random,Toth2018Quantum}.
New approaches have been found to obtain the quantum Fisher information in systems in thermal equilibrium by measuring certain observables \cite{Hauke2016Measuring,Shitara2016Determining}. Using the theory of quantum Fisher information, it has  been examined how the precision  scales with the size of a noisy quantum system \cite{Escher2011General,Demkowicz-Dobrzanski2012The},  which will help to identify cases when very high precision can be achieved with large systems \cite{Chaves2013Noisy}. Connected to these questions, new uncertainty relations have been derived with the  quantum Fisher information, which improve the Heisenberg uncertainty \cite{Frowis2015Tighter}. Moreover, new relations between the quantum Fisher information and the entropy have been presented \cite{Huber2017Geometric,Rouze2017Contractivity}.

Recently, a surprising property of the quantum Fisher information has been discovered: it is, up to a constant factor, the convex roof of the variance \cite{Toth2013Extremal,yu2013quantum}. This result connects the theory of the quantum Fisher information to  entanglement measures, which are also defined by convex roofs \cite{Toth2014Quantum}. The findings of  \REF{Toth2013Extremal,yu2013quantum}  were used to sharpen statements concerning the continuity of the quantum Fisher information \cite{Augusiak2016Asymptotic}.  As another consequence, the quantum Fisher information can efficiently be bounded from below based on few measurements \cite{Apellaniz2017Optimal}.
Finally, the definition of the quantum Fisher information as a convex roof could be used to study the role of entanglement and quantum correlations in interferometry \cite{Bromley2017There}.

In this paper, we present a new approach to bound the quantum Fisher information with the variance and various entropies. In order to list our main results in detail, we need the following definition.

{\bf Definition 1.} We introduce the following quantity
\be
V(\varrho,A):=(\Delta A)^2-\tfrac{1}{4}F_{\rm Q}[\varrho,A].
\label{eq:var_minus_FQover4_simple}
\ee

It is well known that $V(\varrho,A)=0$ for pure states. We can also expect that for states sufficiently pure $V(\varrho,A)$ is small, while for states that are far from pure, the difference can be larger.

We will present methods to bound $V(\varrho,A)$ from above. We now list the three Observations proven in the paper, the proofs will be given later.

{\bf Observation 1.} For rank-2 states $\varrho,$ the difference between the variance and the quantum Fisher information can be obtained  with the purity as
\be
V(\varrho,A)= \tfrac{1}{2}[1-\trace(\varrho^2)](\omega_1-\omega_2)^2,
\label{eq:lambda12}
\ee 
where $\omega_k$ are the eigenvalues of the $2\times 2$ matrix 
\be
\Omega_{kl}=\langle k \vert A \vert l \rangle\label{eq:Akl}
\ee
for $k,l=1,2.$
Here $\ket{1}$ and $\ket{2}$ are the two eigenvectors of $\varrho$ with nonzero eigenvalues. 

Remarkably, \EQ{eq:lambda12} is not an inequality, but an equality, and it connects the quantum Fisher information and the variance to the linear entropy given as
\be
S_{\rm lin}(\varrho)=1-\trace(\varrho^2)=1-\sum_k \lambda_k^2=\sum_{k\ne l} \lambda_k \lambda_l.
\ee

Next, we will present a relation with the linear entropy for density matrices with an arbitrary rank.

\begin{figure}
\centerline{ 
\epsfxsize7.1cm \epsffile{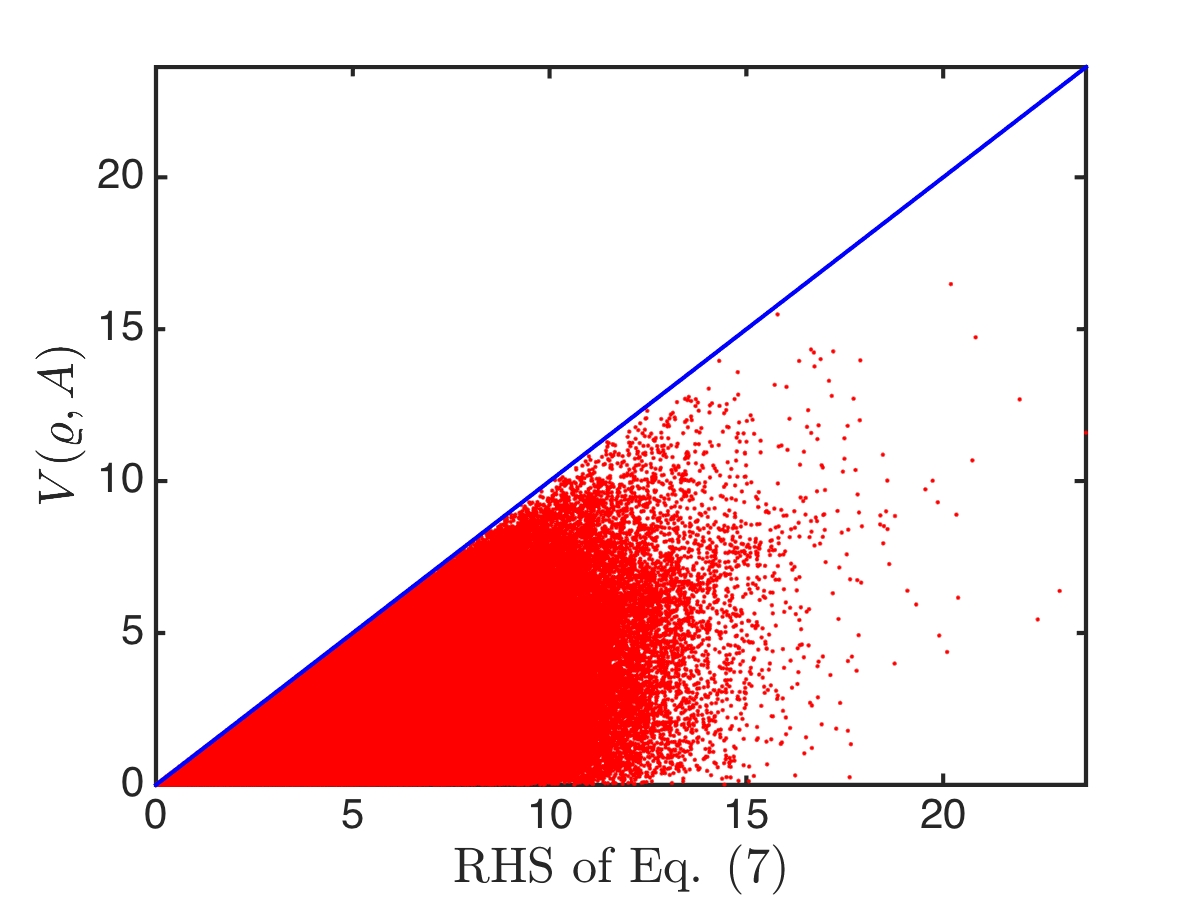}}
\caption{Numerical verification of the inequality \eqref{eq:lambda12bb} for $d=3.$ 
Points corresponding to 10 million random states are shown. For each random state $\varrho,$ a random Hermitian operator $A$ has also been generated. 
All points below the blue line correspond to states that satisfy the inequality. RHS refers to the right-hand side of \EQ{eq:lambda12bb}. The point $(0,0)$ corresponds to pure states. }
\label{fig:numerical_verificiation}
\end{figure}

{\bf Observation 2.} For states $\varrho$ with an arbitrary rank we have
\be
V(\varrho,A)\le \tfrac{1}{2}S_{\rm lin}(\varrho)\left[\sigma_{\max}(A)-\sigma_{\min}(A)\right]^2,
\label{eq:lambda12bb}
\ee 
where $\sigma_{\max}(A)$ and $\sigma_{\min}(A)$ are the largest and smallest eigenvalues, respectively, of 
$A.$ In \FIG{fig:numerical_verificiation}, a numerical verification of the inequality can be seen.

In quantum metrological problems, $A$ is often a collective angular momentum component defined as
\be
J_l=\tfrac{1}{2}\sum_{n=1}^N \sigma_l^{(n)},
\ee
where $l=x,y,z,$ and $\sigma_l^{(n)}$ are the Pauli spin matrices for spin $(n).$ This corresponds to quantum metrology with linear interferometers, which is the most relevant type of metrology with large particle ensembles. For this case, we have to subsititute $\sigma_{\max}(J_l^2)=N^2/4$ into \EQ{eq:lambda12bb}.

So far we considered bounds on $V(\varrho,A)$ for a particular $A.$
We now arrived at computing the average of the quantity \eqref{eq:var_minus_FQover4_simple} for traceless Hermitian operators.

{\bf Observation 3.} The average of $V$ over traceless Hermitian matrices 
is given as
\be
\AVGVSHORT{A} =\frac{2}{d^2-1}\bigg[ S_{\rm lin}(\varrho)+H(\varrho)-1\bigg],\label{eq:avgD}
\ee
where the averaging is over matrices $A$ with a fixed norm ${\rm Tr}(A^2)=2,$ $d$ is the dimension of the system, and 
\be
H(\varrho)=2\sum_{k,l}\frac{\lambda_{k}\lambda_{l}}{\lambda_{k}+\lambda_{l}}=1+2\sum_{k\ne l}\frac{\lambda_{k}\lambda_{l}}{\lambda_{k}+\lambda_{l}}
\label{eq:Hdef}
\ee
is the sum of the pairwise harmonic means of the eigenvalues of $\varrho.$ 
 The quantity $H(\varrho),$ in a certain sense, measures the purity of the quantum state, since for pure states $H(\varrho)=1,$ while for mixed states $H(\varrho)>1.$ 

Our paper is organized as follows. In \SEC{sec:QFIvar}, we present the basics of quantum metrology relevant to our paper, and we show that $V(\varrho,A)$ is a generalized variance defined in \REFS{Petz2002Covariance,Gibilisco2009Quantum}. In \SEC{sec:boundsondifff}, we obtain upper bounds on the the difference between the variance and the quantum Fisher information over four given in \EQ{eq:var_minus_FQover4_simple}. In \SEC{sec:examples}, we show some concrete examples for the application of our inequalities. In \SEC{sec:aver_op}, we calculate averages of the quantum Fisher information and $V(\varrho,A)$ over all Hermitian operators, and relate them to the von Neumann entropy.
Finally, we calculate similar averages for the Kubo-Mori-Bogoliubov quantum Fisher information.

\section{Basics of quantum metrology}

\label{sec:QFIvar}

\subsection{The quantum Fisher information and the variance}

In this section, we review important properties of the quantum Fisher information and the variance. We will also stress the relations that connect these two quantities,
which will motivate us to study $V(\varrho,A)$ in the rest of the paper.

The quantum Fisher information can be computed as follows.
Let us assume that  a density matrix is given in its eigenbasis as 
\begin{equation}
\varrho=\sum_{k=1}^d \lambda_k \ketbra{k},
\end{equation}
where $d$ is the dimension of the quantum system.
Then, the quantum Fisher information is obtained as \cite{Helstrom1976Quantum,Holevo1982Probabilistic,Braunstein1994Statistical,Braunstein1996Generalized}
\begin{equation}
F_{\rm Q}[\varrho,A]=2\sum_{k,l}\frac{(\lambda_{k}-\lambda_{l})^{2}}{\lambda_{k}+\lambda_{l}}\vert A_{kl} \vert^{2},\label{eq:FQ}
\end{equation}
where $A_{kl}$ is defined as $A_{kl}=\langle k \vert A \vert l \rangle.$
\EQL{eq:FQ} can be rewritten as \cite{Toth2014Quantum}
\begin{eqnarray}
F_{\rm Q}[\varrho,A]&=&4\sum_{k,l}\lambda_{k}\vert A_{kl}  \vert^{2}-8\sum_{k,l}\frac{\lambda_{k}\lambda_{l}}{\lambda_{k}+\lambda_{l}}\vert A_{kl} \vert^{2}\nonumber\\
&=&4\exs{A^2}-8\sum_{k,l}\frac{\lambda_{k}\lambda_{l}}{\lambda_{k}+\lambda_{l}}\vert  A_{kl}  \vert^{2}.
\label{eq:qF2}
\end{eqnarray}
The advantage of \EQ{eq:qF2} is that the $\exs{A^2}$ appears in the formula, which
makes it easy to compare the quantum Fisher information to the 
the variance given as
\begin{eqnarray}
(\Delta A)^2&=&\exs{A^2}-\ex{A}^2\nonumber\\
&=&\sum_{k,l} \lambda_k \vert A_{kl} \vert ^2-\left(\sum_k \lambda_k A_{kk} \right)^2.
\label{eq:var}
\end{eqnarray}

For any decomposition $\{p_{k},\ket{\Psi_{k}}\}$ of the density matrix $\varrho$ we have \cite{Toth2013Extremal,yu2013quantum}
\begin{equation}\label{eq:boundavF}
\tfrac{1}{4}F_{\rm Q}[\varrho,A]  \le  \sum_k p_k \va {A}_{\Psi_k} \le \va A_{\varrho},
\end{equation}
where the upper and the lower bounds are both tight in the sense that there are 
decompositions that saturate the first inequality, and there are others that saturate the second one. Note that the latter statement could be generalized to covariance matrices \cite{Petz2013A,Leka2013Some}.

These statements can also be expressed saying that the quantum Fisher information over four is the convex roof of the variance
\be
\frac{1}{4}F_{\rm Q}[\varrho,A]=\inf_{\{p_k,\Psi_k\}}\sum_k p_k (\Delta A)_{\Psi_k}^2,
\ee 
while the variance is the concave roof of the itself
\be
(\Delta A)_{\varrho}^2=\sup_{\{p_k,\Psi_k\}}\sum_k p_k (\Delta A)_{\Psi_k}^2,
\ee 
where the infimum and the supremum are over all possible convex decompositions of $\varrho$ of the type 
\be
\varrho=\sum_k p_k\ketbra{\Psi_k}.\label{eq:decomp_nonorth} 
\ee
where $p_k$ are probabilities and $\ket{\Psi_k}$ are pure states.
Finally, we note that we can also interpret the relation of the quantum Fisher information and the variance as follows. We write the variance as
\cite{Toth2013Extremal}
\be
\va A=\sum_{k}p_{k}\va{A}_k+\sum_{k}p_{k}(\exs A-\exs A_{k})^{2}.\label{eq:vasplit}
\ee
\EQL{eq:vasplit} is valid for all decompositions of $\varrho$ of the type \eqref{eq:decomp_nonorth}. The first term on the right-hand side of \EQ{eq:vasplit} we can call "quantum" part, since it comes from the variance of the operator on pure quantum states \cite{Toth2013Extremal}. The second term we can call the "classical" part, since it is just a classical variance formula for the subensemble expectation values \cite{[{Another approach for defining classical  and quantum concerning the quantum Fisher information can be found in }] [{.}]Alipour2015Extended}. In this picture, we can interpret the quantum Fisher information as the minimal "quantum" part of the variance, while $V(\varrho,A)$ given in \EQ{eq:var_minus_FQover4_simple} is the maximum of the "classical" part. Hence, we can define $V(\varrho,A)$ as a concave roof as 
\be\label{eq:Vasconcaveroof}
V(\varrho,A)=\sup_{\{p_k,\Psi_k\}}\sum_k p_k (\exs A_{\Psi_k}-\exs A)^{2},
\ee
which can also be rewritten as 
\be\label{eq:Vasconcaveroof2}
V(\varrho,A)=\sup_{\{p_k,\Psi_k\}}\sum_k p_k (\exs A_{\Psi_k}^{2}-\exs{A}^2).
\ee

\subsection{The difference between the variance and the quantum Fisher information over four}
\label{eq:difference}

In this section, we discuss some important properties of $V(\varrho,A).$ We also discuss that
$V(\varrho,A)$ equals a generalized variance given in \REFS{Petz2002Covariance,Gibilisco2009Quantum}.

The quantity \eqref{eq:var_minus_FQover4_simple} has the following important properties.

\begin{itemize}
\item[(i)] $V(\varrho,A)=0$ for all pure states, and for all states for which  
$(\Delta A)^2=0.$
\item[(ii)] $V(\varrho,A)=0$ for all $A$ if and only if $\varrho$ is pure.
\item[(iii)] Since the variance is concave in the state and the quantum Fisher
information is convex, it is also concave in the state
\item[(iv)] The quantity $V(\varrho,A)$ is clearly independent from $\trace(A),$ that is, 
$V(\varrho,A)=V(\varrho,A+c\openone)$ for any $c.$
\item[(v)] The relation $V(U\varrho U^\dagger,A)=V(\varrho,U^\dagger A U)$ holds for any unitary $U.$ 
\item[(vi)] For a bipartite system, for product states 
$V(\varrho_A\otimes\varrho_B,\openone \otimes A + B \otimes \openone)=V(\varrho_A,A)+V(\varrho_B,B)$ holds.
This can be proven noting that a similar relation holds for the variance and the quantum Fisher information.

\end{itemize}

Next, we present some formulas defining $V(\varrho,A).$ From \EQ{eq:qF2} and \EQ{eq:var}, we obtain
\begin{eqnarray}
V(\varrho,A)=
2\sum_{k,l}\frac{\lambda_{k}\lambda_{l}}{\lambda_{k}+\lambda_{l}} \vert A_{kl}  \vert^{2}
-\left(\sum\lambda_{i}A_{ii}^{\text{}}\right)^{2}.
\label{eq:var_minus_FQover4}
\end{eqnarray}
It is instructive to regroup the terms such that the double sum is only over indices that are not equal with each other. Hence, an alternative form of \EQ{eq:var_minus_FQover4} is obtained as
\begin{eqnarray}
V(\varrho,A)&=&
2\sum_{k\ne l}\frac{\lambda_{k}\lambda_{l}}{\lambda_{k}+\lambda_{l}} \vert A_{kl}  \vert^{2}\nonumber\\
&+&\sum_k \lambda_k A_{kk}^2  -\left(\sum\lambda_{k}A_{kk}^{\text{}}\right)^{2}.
\label{eq:var_minus_FQover4_alt}
\end{eqnarray}
One is tempted to think that the last two terms in \EQ{eq:var_minus_FQover4_alt} is the variance $\va{A}.$ Indeed, they are equal to the variance if $A$ is diagonal in the eigenbasis of $\varrho.$ Otherwise, one can realize that the equality does not hold in the general case by comparing these terms to \EQ{eq:var}.

It is instructive to connect $V(\varrho,A)$ to the family of generalized variances
defined in \REFS{Petz2002Covariance,Gibilisco2009Quantum} as
\begin{eqnarray}
{\rm var}_{\varrho}^{f}(A)&=&\sum_{ij}m_{f}(\lambda_{i},\lambda_{j})\vert A_{ij} \vert^2
-\left(\sum\lambda_{i}A_{ii}^{\text{}}\right)^{2},
\label{eq:qCov}
\end{eqnarray}
where $f:\mathbbm{R}^+\rightarrow\mathbbm{R}^+$ is a matrix monotone function, and 
\be 
m_f(a,b)=bf(a/b)
\label{eq:mean}
\ee 
is a corresponding mean. With \EQ{eq:qCov} we can define a large set of generalized variances. 
The $f(x)$ are boudned as
\be
f_{\min}(x)\le f(x) \le f_{\max}(x),
\ee
where the lower and upper bounds, respectively, are given as
\be
f_{\min}(x)=\frac{2x}{1+x}, \quad f_{\max}(x)=\frac{1+x}{2}.
\ee
The generalized variance
\eqref{eq:qCov} with $f(x)=f_{\max}(x)$ is the usual variance 
\be
{\rm var}^{\max}_\varrho(A)=\ex{A^2}-\ex{A}^2,
\ee
where in the subscript there is "$\max$" rather than $f_{\max}$ for simplicity.
The corresponding mean is the arithmetic mean  $m_{\max}(a,b)=(a+b)/2.$
Let us now consider the generalized variance
\eqref{eq:qCov} with $f_{\min}(x).$  The corresponding mean is the harmonic mean $m_{\min}(a,b)=2ab/(a+b).$  Straightforward calculations show that
\be
{\rm var}^{\min}_\varrho(A)\equiv V(\varrho,A)
\ee
holds
\cite{*[{Another type of generalized variance appears in }] [{, where the quantum variance is defined as ${\rm var}_{\rm Q}=\va{A}-{\rm var}^{f_{\rm log}}_\varrho(A),$ where $f_{\log}(x)=(x-1)/\log(x)$ and the corresponding mean is the logarithmic mean $m_{\log}(a,b)=(b-a)/(\log(b)-\log(a)).$  The quantity ${\rm var}^{f_{\rm log}}_\varrho(A)$ is also called the Kubo-Mori-Bogoliubov variance \cite{Petz2002Covariance,Gibilisco2009Quantum}. 
It is connected to the magnetic susceptibility $\kappa_z=\partial\exs{J_z}/\partial B$ in systems in thermal equlibrium as $\partial{\kappa_z}/\partial{B}={\rm var}^{f_{\rm log}}_\varrho(J_z)/kT$. Here $B$ is the magnetic field, $T$ is the temperature and $k$ is the Boltzmann constant. 
\phantom{\cite{Nolan2017Quantum}}
}] FrerotQuantum2016}.
Clearly, 
${\rm var}^{\min}_\varrho(A)$ is the smallest of the generalized variances, while the usual variance, ${\rm var}^{\max}_\varrho(A)$, is the largest \cite{Petz2002Covariance,Gibilisco2009Quantum}. For any generalized variance given in \EQ{eq:qCov} we have
\be
\va{A}-F_{\rm Q}[\varrho,A]\le {\rm var}^{f}_\varrho(A)\le \va{A}.
\ee
Hence, if $F_{\rm Q}[\varrho,A]=0$ then all the generalized variances give the same value.

\section{Upper bound on $V(\varrho)$ with the purity}
\label{sec:boundsondifff}

In this section, we prove Observations 1-2.

{\it Proof of Observation 1.} For the rank-$2$ case, $\lambda_1=\lambda$ and $\lambda_2=1-\lambda,$ and all other eigenvalues are zero. Then, \EQ{eq:var_minus_FQover4}
becomes
\be
V(\varrho,A)=\lambda(1-\lambda)\left[(A_{11}-A_{22})^2+4\vert A_{12}\vert^2\right].
\label{eq:A0011b}
\ee
Now, first we have to use the following relation 
\be
\lambda(1-\lambda)=\tfrac{1}{2}\left[1-\trace(\varrho^2)\right],
\ee
which can easily be proved with direct calculation.
Second, we have to show that
\be
(A_{11}-A_{22})^2+4\vert A_{12}\vert^2=(\omega_1-\omega_2)^2,\label{eq:eigenvalues}
\ee
where $w_{1,2}$ are the eigenvalues of the matrix
\be
\left(
\begin{array}{cc}
  A_{11} &  A_{12}   \\
  A_{12}^* &   A_{22}  \\
\end{array}
\right).\label{eq:A0011def}
\ee
This can be seen using the usual formula for the eigenvalues of a $2\times 2$ Hermitian matrix given as
\be
\omega_{1,2}=\frac{(A_{11}+A_{22})\pm\sqrt{D}}{2},\label{eq:omega12}
\ee
where
\be
D=(A_{11}+A_{22})^2-4(A_{11}A_{22}-\vert A_{12}\vert^2).\label{eq:D}
\ee
Equations~\eqref{eq:omega12} and \eqref{eq:D} yield \EQ{eq:eigenvalues}.
With these, we have proved the equality, \eqref{eq:lambda12}, giving $V(\varrho,A)$ as a function of the purity and the eigenvalues. $\qed$

So far, we found an upper bound on $V(\varrho,A)$ for states with rank at most two. 
Next, we will look for a bound for states with an arbitrary rank, in order to prove Observation 2.

{\it Proof of Observation 2.} Based on the definition of $V(\varrho,A)$ as a convex roof given in \EQ{eq:Vasconcaveroof}, we see that the relation \eqref{eq:lambda12bb} is true, if and only if
\bea
&&\tfrac{1}{2}S_{\rm lin}\left(\sum_k p_k\ketbra{\Psi_k}\right)\left[\sigma_{\max}(A)-\sigma_{\min}(A)\right]^2\nonumber\\
&&\quad\quad-\sum_k p_k \left(\exs A_{\Psi_k}-\exs A\right)^{2}\label{eq:Vasconcaveroof2}
\eea
is non-negative for all possible choices for $p_k$ and $\ket{\Psi_k}.$ In order to verify that \EQ{eq:Vasconcaveroof2} cannot be negative, we need to minimize  it over $p_k$ and $\ket{\Psi_k}.$ Let us consider now only a minimization over $\vec p=(p_1,p_2,p_3,...)$ under the constraints $p_k\ge 0$, $\sum_k p_k=1.$ We consider a further constraint for the expectation value, $\ex A=\sum_k p_k \exs A_{\Psi_k}=A_0,$ where $A_0$ is a constant.  While we minimize over $\vec p,$ we keep the $\ket{\Psi_k}$ fixed.  

Next, we will determine the characteristics of the $\vec p$\;'s that minimize \EQ{eq:Vasconcaveroof2}. The first term in \EQ{eq:Vasconcaveroof2} is concave in the $p_k$'s, the second term is linear. Then, \EQ{eq:Vasconcaveroof2} is also a concave function of  $p_k$'s, and takes its minimum on the extreme points of the convex set of the allowed values for $\vec p.$ The extreme points correspond to cases where at most two of the $p_k$'s are non-zero. (If we did not have the $\sum_k p_k \exs A_{\Psi_k}=A_0$ constraint, then the extreme points would correspond to $\vec p$\;'s with at most a single nonzero $p_k.$)
Thus, we need to prove that \EQ{eq:Vasconcaveroof2} is non-negative for such cases. This can straightforwardly be done based on   \EQ{eq:lambda12}, which gives $V(\varrho,A)$ for rank-$2$ states with the purity. 

We have just proved that for any choice of $\ket{\Psi_k}$, minimzing over $\vec p$ will lead to a nonnegative value for \EQ{eq:Vasconcaveroof2}. From this, the statement of the observation follows.
$\qed$

\section{Examples}
\label{sec:examples}

Next, we examine how our lower bounds on the quantum Fisher information behave in some relevant situations.

\subsection{Pure states}

As a warm-up excercise, let us consider pure states.
For a pure state $\ket{\Psi}$ the relations
\bea
V(\ket{\Psi},A)&=&0,\nonumber\\
S_{\rm lin}(\ket{\Psi})&=&0,\nonumber\\
H(\ket{\Psi})&=&1
\eea
hold for any $A.$ Clearly, pure  states saturate \EQ{eq:lambda12bb}.

\subsection{Completely mixed state}

The completely mixed state is defined as
\begin{equation}\label{eq:rhocm}
\varrho_{\rm cm}=\frac{\openone}{d},
\end{equation}
where $d$ is the dimension of the system. The state $\varrho_{\rm cm}$ is not useful for metrology since $F_{\rm Q}[\varrho,A]=0$ for all $A.$ In fact, $\varrho_{\rm cm}$ is the only quantum state that has this property. Hence, $V(\varrho_{\rm cm},A)$ equals the variance of the state
\begin{equation}
V(\varrho_{\rm cm},A)=\va{A}_{\varrho_{\rm cm}}=\frac{1}{d}\trace(A^2),
\end{equation}
where we assumed that $A$ is traceless.
The linear entropy is maximal for the completely mixed state
\begin{equation}
S_{\rm lin}(\varrho_{\rm cm}^2)=1-\frac{1}{d}.
\end{equation}
Finally, $H$ defined in \EQ{eq:Hdef} is also maximal 
\begin{equation}\label{eq:rhocmH}
H(\varrho_{\rm cm})=d.
\end{equation}

Let us see, whether $\varrho_{\rm cm}$ saturates \EQ{eq:lambda12bb}. Let us consider $N$ qubits corresponding to $d=2^N.$ Direct calculations shows that the completely mixed state saturates \EQ{eq:lambda12bb} only for the $d=2$ case, i.e., for a single qubit.

\subsection{GHZ states}

In this section, we consider states that live in the two-dimensional subspace  
\be\{\ket{000..00},\ket{111..11}\}.\label{eq:2dspace}\ee 
Such states  are very relevant for experiments with trapped ions aiming to create GHZ states \cite{Leibfried2004Toward,Sackett2000Experimental,Monz201114-Qubit}
defined as
\be
\ket{\rm GHZ}=\tfrac{1}{\sqrt{2}}\left( \ket{000..00}+\ket{111..11}\right).
\ee
For states of the type \be\label{eq:noisyGHZ}
\varrho_p=p\frac{P_{000..00}+P_{111..111}}{2}+(1-p)\ketbra{{\rm GHZ}}
\ee
a relation giving the quantum Fisher information as a function of the density matrix
\be\label{eq:FQGHZ}
F_{\rm Q}[\varrho,J_l]= 2N^2\left[\trace(\varrho^2)-1\right]
\ee
holds \cite{Nolan2017Quantum}. \EQL{eq:FQGHZ} has also been found in the context of relating the visibility to the metrological performance in ion-trap experiments in \REF{Pezze2016Witnessing}.

Let us apply now our theory to obtain a bound for any state living 
in the space \eqref{eq:2dspace}.
Such states satisfy \EQ{eq:lambda12} with $\omega_1=N/2$ and 
$\omega_2=-N/2.$ Simple algebra shows that 
all such states saturate \EQ{eq:lambda12bb} with $A=J_z.$
For states of the two-dimensional subspace given in \eqref{eq:2dspace}, the variance of $J_z$ is given as 
\be\label{eq:varGHZ}
\va{J_z}=\left(1-\exs{P_{000..00}}^2-\exs{P_{111..11}}^2\right)\frac{N^2}{2}.
\ee
Using that \EQ{eq:lambda12bb} is saturated and from \EQ{eq:varGHZ} we obtain
\be
\frac{F_{\rm Q}[\varrho,J_z]}{N^2}= 2\left[\trace(\varrho^2)-\exs{P_{000..00}}^2-\exs{P_{111..11}}^2\right].\label{eq:rank2bound_general2}
\ee 
For noisy GHZ states of the form \eqref{eq:noisyGHZ}, the relation \eqref{eq:rank2bound_general2} reduces to \EQ{eq:FQGHZ}.

We mention that another lower bound on the quantum Fisher information with the fidelity $F_{\rm GHZ}$ is given by \cite{Apellaniz2017Optimal}
\be\label{eq:FQFGHZ}
\frac{F_{\rm Q}[\varrho,J_z]}{N^2}\ge \bigg\{\begin{array}{ll}
 (1-2F_{\rm GHZ})^2, & \text{ if } F_{\rm GHZ}> 1/2,\\
0,& \text{ if } F_{\rm GHZ}\le 1/2.
\end{array}
\ee
The bound \eqref{eq:FQFGHZ} is valid for any quantum state, even for the ones that do not live in the two-dimensional subspace. The states within the two-dimensional subspace do not all saturate \EQ{eq:FQFGHZ}.

\subsection{Ensemble of spin-$\frac{1}{2}$ particles}

In this section, we apply our bound \EQ{eq:lambda12bb} to an ensemble of spin-$\frac{1}{2}$ particles.

It has been shown that for separable states in a linear interferometer, the quantum Fisher information is bounded as \cite{Pezze2009Entanglement}
\be\label{eq:FQN}
F_{\rm Q}[\varrho,J_l]\le N.
\ee
Any state that violates \EQ{eq:FQN} is entangled. For general states the bound is
\be
F_{\rm Q}[\varrho,J_l]\le N^2,
\ee
which is called the Heisenberg-limit. 

It is an important question in metrology, what the conditions are for the Heisenberg scaling given by 
\be
F_{\rm Q}[\varrho,J_l]= O(N^2),
\ee
where $O$ is the usual Landau symbol. Rewriting \EQ{eq:lambda12bb} for $A=J_z,$ we arrive at
\be
\va{J_l}-\tfrac{1}{4}F_{\rm Q}[\varrho,J_l]\le \left[1-\trace(\varrho^2)\right]\frac{N^2}{2}.
\ee 
Let us now consider a family of states $\varrho_N$ such that $S_{\rm lin}(\varrho_N^2)\le s,$ where $s$ is some constant.
Then, if 
\be
\va{J_l}_{\varrho_N}\ge s\frac{N^2}{2},
\ee
then we have Heisenberg scaling. That is, it is sufficient that the variance scales as $O(N^2)$ and the state is sufficiently pure.

\subsection{Systems in thermal equalibrium}

If our bounds are used in systems in thermal equilibrium 
then the purity $\trace(\varrho^2)$ can straightforwardly be obtained from the temperature, using that the eigenvalues of the density matrix are given as 
\be
\lambda_l \propto e^{\frac{E_l}{k_{\rm B}T}},
\ee
where $E_k$ are the energy levels of the system, $T$ is the temperature and $k_{\rm B}$ is the Boltzmann constant. Hence, using Observation 2, we can bound the quantum Fisher information from below if we know the variance and the temperature of the system. The method gives a useful bound if $kT\lesssim E_1-E_0.$

\section{Averaging over operators}
\label{sec:aver_op}

In this section, we determine the averages over all operators for the variance, the quantum Fisher information and $V(\varrho,A).$ This sheds new light on the relation between these quantities and entropies.

In \SEC{eq:difference} it has been discussed that  $V(\varrho,A)=0$ for all  $A$ if and only if the state $\varrho$ is pure.  If we then average $V(\varrho,A)$ over all observables $A,$ we obtain a quantity that is zero only for pure states. This quantity is concave in $\varrho,$ since $V(\varrho,A)$ is also concave. Hence, it seems to be interesting to ask, how it is related to entropies.

\subsection{Averaging over the Hermitian matrices}
\label{sec:av_variance}

Next, we will discuss how to interpret the averaging over all traceless Hermitian matrices with a given norm.
All such matrices can be obtained as a linear combination of the SU($d$) generators as
\be
A_{\vec n}:=\vec{A}^T \vec{n},\label{eq:An}
\ee
where $\vec{A}=[A^{(1)},A^{(2)},A^{(3)},...]^T,$  $\vec{n}$ is a unitvector with real elements, and $(.)^T$ denotes matrix transpose. 
We consider the following normalization
\be
{\rm Tr}(A^{(k)}A^{(l)})=2\delta_{kl}.\label{eq:trAkAl}
\ee
According to well-known results of linear algerbra, the number of SU($d$) generators is  
\be 
N_{\rm g}=d^2-1.
\ee 

We now define the average over unit vectors as
\be
\int f(\vec n) M(d\vec n),
\ee
where $M$ is a measure over unitvectors with the usual invariance properties such that $\int M(d\vec n)=1,$ and $f(\vec n)$ is some function depending on the unit vector $\vec n.$ Hence, we can average an expression over traceless Hermitian matrices with a given norm as 
\be
\AVGfA=\int f(A_{\vec n}) M(d\vec n).
\ee

Let us now calculate the average of the variance over the Hermitian matrices.
For any operator $A_{\vec n}$ one can obtain the variance as
\be\label{eq:varAn}
\va{A_{\vec n}}=\vec{n}^T C \vec{n},
\ee
where $C$ is the covariance matrix defined as 
\be C_{mn}=\tfrac{1}{2}\left(\ex{A^{(m)} A^{(n)}}+\ex{A^{(n)} A^{(m)}}\right)-\ex{A^{(m)}}\exs{A^{(n)}}.\ee
Then, the average variance can be written 
as a sum of the variances of the generators $A^{(k)},$ since
\bea
&&\AVGVAR{A}\nonumber\\
&&\quad\quad=
\int M(d\vec n) \; \vec{n}^T C \vec{n}
=\int M(d\vec n) \;  \trace(C \vec{n} \vec{n}^T)\nonumber\\
&&\quad\quad= \frac{1}{N_{\rm g}} \trace\left(C\right)=
\frac{1}{N_{\rm g}}\sum_{m=1}^{N_{\rm g}} \va{A^{(m)}},
\label{eq:avgvar}
\eea
where we used that
\be
\int M(d\vec n) \; \vec{n} \vec{n}^T=\frac{\openone}{N_{\rm g}}.
\ee
Based on \EQ{eq:avgvar}, $\AVGVAR{A}$ is independent from the concrete choice of the  $A^{(k)}.$

Following the ideas above also for the quantum Fisher information, we present now relations for the averages of various quantites.

{\bf Observation 4.} For $d\times d$ systems, the averages of the variance and the quantum Fisher information, respectively, are
\begin{subequations}
\bea\label{eq:sumvar}
\AVGVAR{A}&=&\frac{2}{N_{\rm g}} \left[ S_{\rm lin}(\varrho)+d-1 \right],\\\label{eq:sumFQ}
\AVGFQ{A}&=&\frac{8}{N_{\rm g}} [d-H(\rho)].
\eea
\end{subequations}
The averages of the off-diagonal and diagonal elements of $A,$ respectively, used later in calculations are
\begin{subequations}
\bea\label{eq:avgAn12}
\AVGAKK{A}&=&\frac{2}{N_{\rm g}},\\
\label{eq:avgA11}
\AVGAKL{A}&=&\frac{2}{N_{\rm g}}\left(1-\frac{1}{d}\right),
\eea
\end{subequations}
where $k\ne l.$ The proof is given in \APP{app:avF}.

\begin{figure}
\centerline{ 
\epsfxsize4.1cm \epsffile{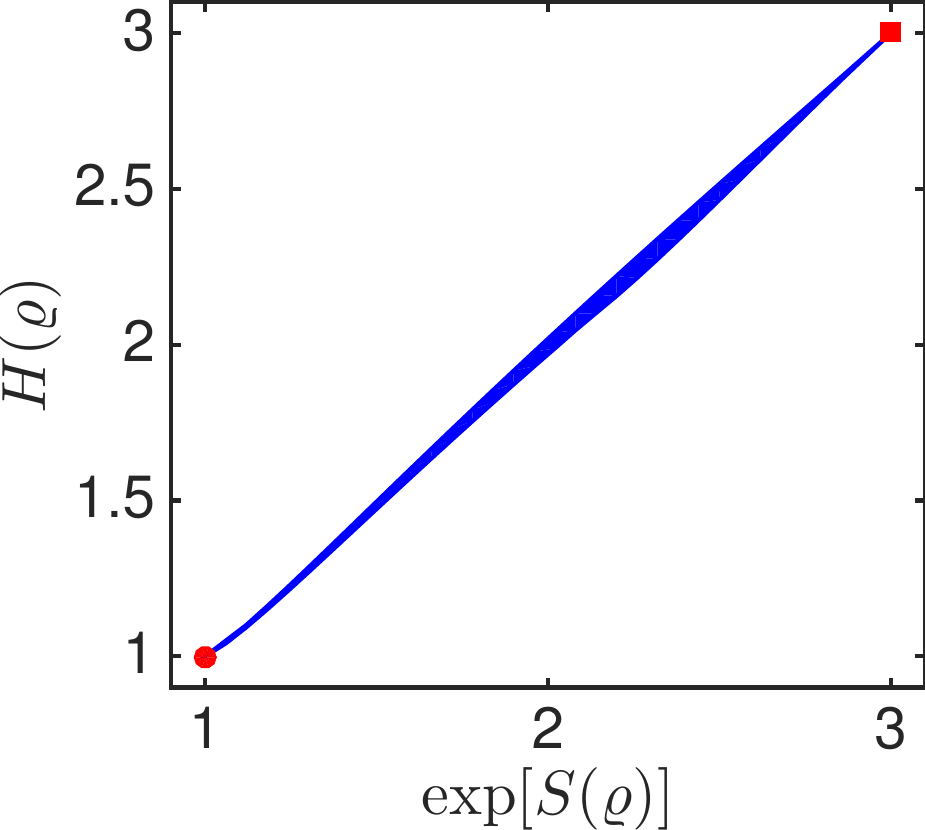}\hskip0.4cm
\epsfxsize4.05cm \epsffile{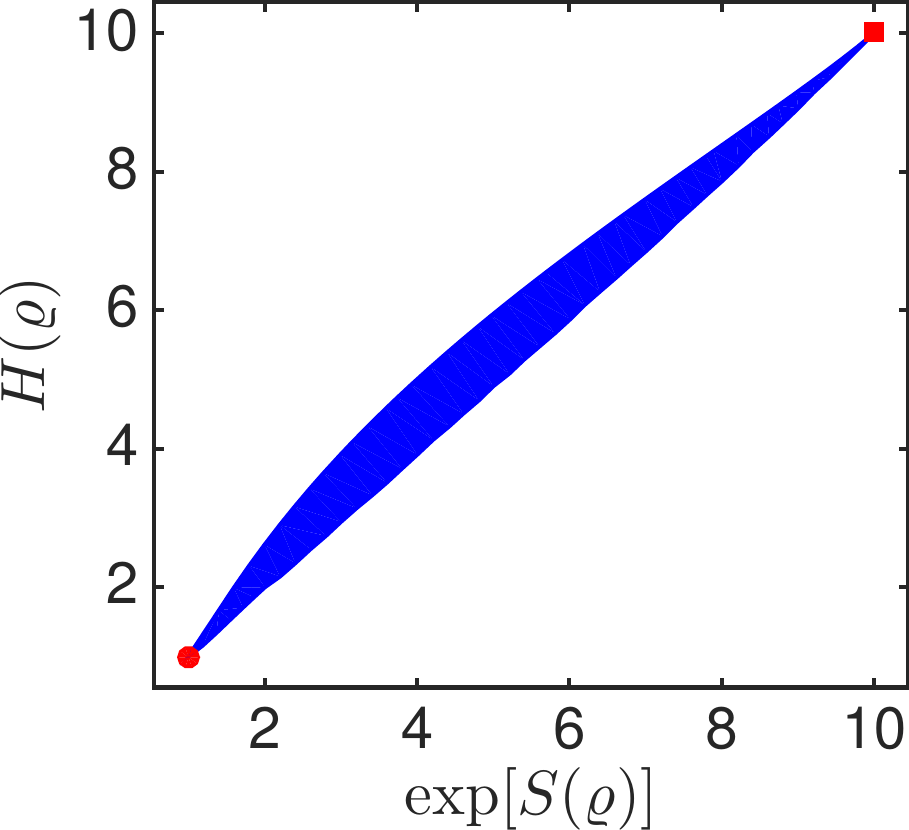}}
\caption{The relation between the von-Neumann entropy and $H(\varrho)$ defined in \EQ{eq:Hdef} for (left) $d=3$ and (right) $10.$  (filled area) Physical quantum states. (circle) Pure states. (square) Completely mixed state \eqref{eq:rhocm}.}
\label{fig:H}
\end{figure}

Simple numerical optimization shows a remarkable relation between $H(\varrho)$ and the von Neumann entropy 
\be \label{eq:defS}
S=-{\rm Tr}(\varrho \log \varrho)=\sum_{k=1}^d \lambda_l \ln \lambda_k,
\ee 
where $\ln(x)$ is the natural logarithm. In \FIG{fig:H}, we indicated the part of the $(H,\exp(S))$-space allowed for physical states. The exponential of the entropy defined as
\be\label{eq:expS}
\exp(S)=\prod_{k=1}^d \lambda_k^{-\lambda_k}
\ee 
has attracted considerable attention \cite{DePalma2014AGeneralization,Konig2014The}.
\FIG{fig:H} also supports the that  $H(\varrho)$ is, essentially, a measure of purity that 
is related to the von Neumann entropy. We find the approximate relation
\be\label{eq:HexpS}
H(\varrho) \sim \exp[S(\varrho)].
\ee
\EQL{eq:HexpS} can be proved for states close to the completely mixed state
based on an expansion of both sides around the point
given by $\lambda_k^{(0)}=1/d$ for $k=1,2,..d.$ 
After we set $\lambda_d=1-\sum_{k=1}^{d-1}\lambda_k,$ 
such an expansion involves $\lambda_k$ for $k=1,2,..,d-1.$
The left-hand side and the right-hand side of \EQ{eq:HexpS} are equal up to second order in the quantities $(\lambda_k-\lambda_k^{(0)}).$  

Based on \EQ{eq:sumFQ}, for the quantum Fisher information we obtain
\be\label{eq:sumFQ2}
\AVGFQ{A}\sim\frac{8}{N_{\rm g}} \left\{d-\exp[S(\varrho)]\right\}.
\ee

Note that the relation \EQ{eq:HexpS} can be used to approximate the von Neumann entropy with a quantity that is easier to compute. 
Note also that our findings are very relevant to recent efforts to obtain inequalities between the quantum Fisher information and the von Neumann entropy \cite{Huber2017Geometric,Rouze2017Contractivity}. We will discuss this in \SEC{sec:FQmath} in more detail.

\begin{figure}
\centerline{ 
\epsfxsize4.1cm \epsffile{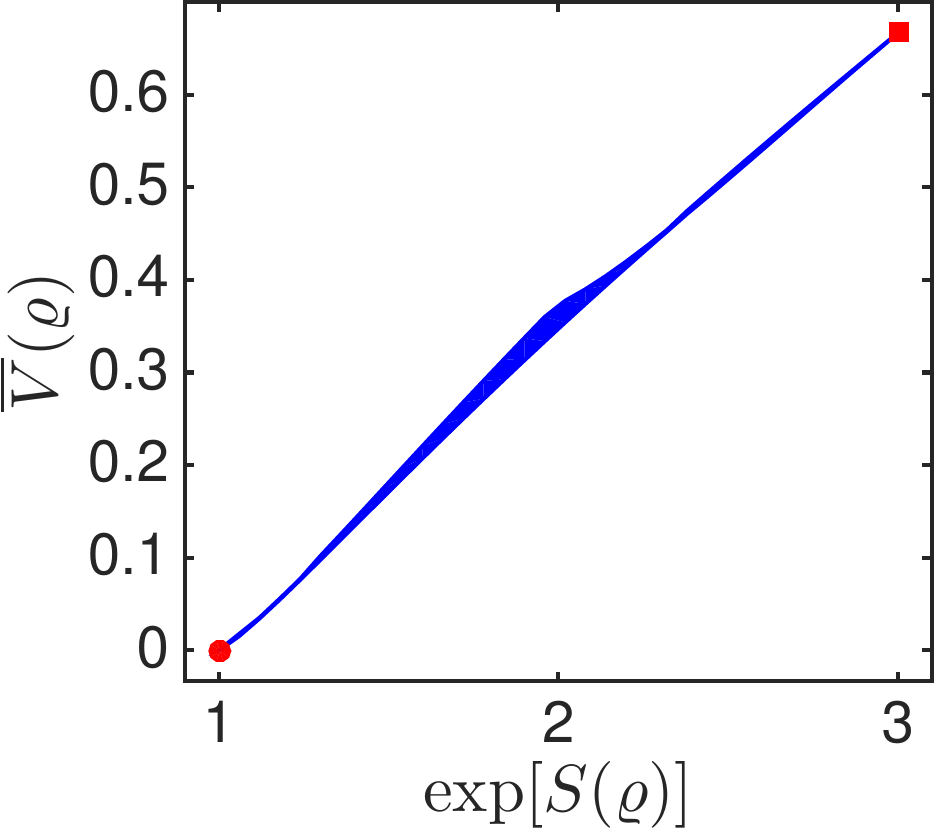}\hskip0.4cm
\epsfxsize4.25cm \epsffile{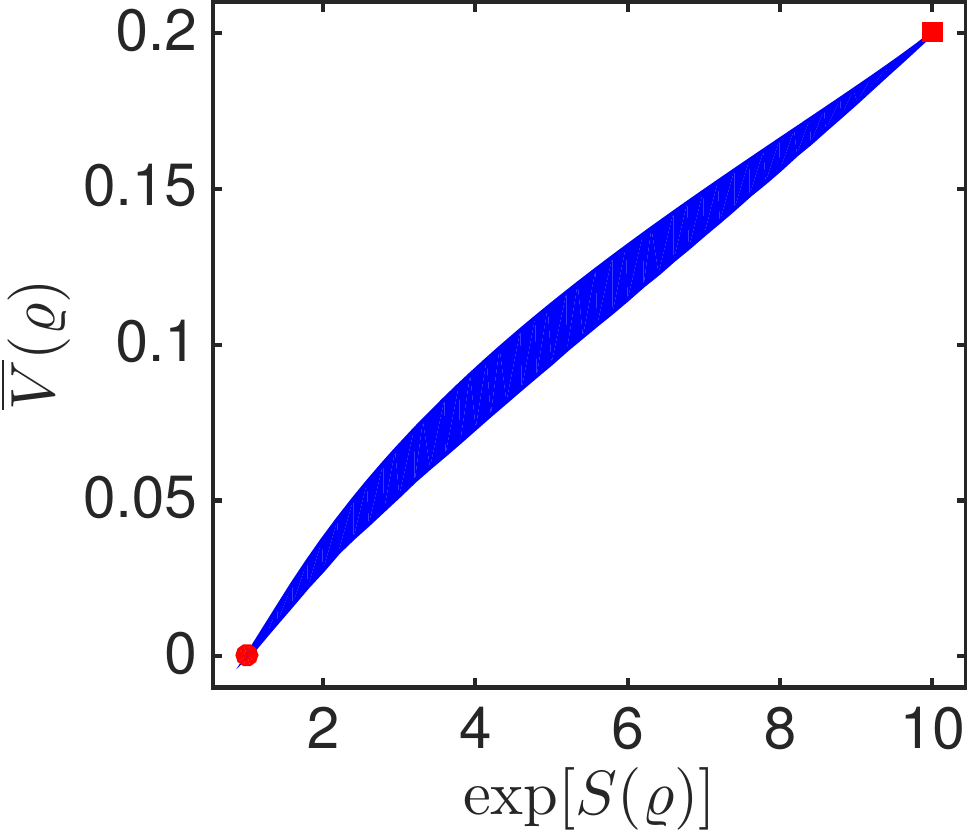}}
\caption{The relation between the von-Neumann entropy  $S(\varrho)$  and the average $V(\varrho)$ defined in \EQ{eq:avgD}  for (left) $d=3$ and (right) $10.$ (filled area) Physical quantum states. (circle) Pure states. (square) Completely mixed state \eqref{eq:rhocm}.}
\label{fig:D}
\end{figure}

\subsection{Averaging $V(\varrho,A)$  over traceless Hermitian operators}

{\it Proof of Observation 3.} Finally, we arrived at computing the average of the quantity \eqref{eq:var_minus_FQover4_simple} for Hermitian operators. We have to use Observation 3 and that
$\AVGV{A}=\AVGVAR{A}-\AVGFQ{A}/4.$
The result is given in \EQ{eq:avgD}.
In \FIG{fig:D}, \EQ{eq:avgD} and $\exp(S)$ are shown for random states of dimension $d=3$ and $d=10.$ Now the correlation with $\exp(S)$ seems to be even more pronounced than in the case of the average quantum Fisher information. We find the approximate relation
\be\label{eq:avgV}
\AVGV{A}\sim \frac{2}{N_{\rm g}} \left(1-\frac{1}{d^2}\right)\exp[S(\varrho)].
\ee

\subsection{The Kubo-Mori-Bogoliubov quantum Fisher information}
\label{sec:FQmath}

We now consider another form of the quantum Fisher information used frequently in mathematics \cite{Petz2008Quantum}, and
calculate its average over traceless Hermitian operators. The  Kubo-Mori-Bogoliubov quantum Fisher information $F_{\rm Q}^{\log}[\varrho,A]$ can be expressed  as
\begin{equation}
F_{\rm Q}^{\log}[\varrho,A]= \sum_{k,l}
\left[\log(\lambda_{k})-\log(\lambda_{l})\right](\lambda_{k}-\lambda_{l})
\vert A_{kl} \vert^{2},\label{eq:FQlog_physics}
\end{equation}
which is related to the relative entropy \be S(\varrho\vert\vert\sigma)={\rm Tr}(\varrho \log \varrho)-{\rm Tr}(\varrho \log \sigma)\ee via the following expression
\begin{equation}\label{eq:FlogSrel}
\frac{d^2}{d^2 \theta}S(\varrho\vert\vert e^{-iA \theta}\varrho e^{+iA \theta})\vert_{ \theta=0}
=F_{\rm Q}^{\log}[\varrho,A].
\end{equation}
[For a derivation of \EQS{eq:FQlog_physics} and \eqref{eq:FlogSrel}, see \APP{app:Srel}.]

Using \EQ{eq:avgAn12}, the average of the quantum Fisher information over the SU(d) generators is given as
\begin{equation}
\AVGFQLOG{A}=\frac{2}{N_{\rm g}}
\sum_{k,l}\left[\log(\lambda_{k})-\log(\lambda_{l})\right](\lambda_{k}-\lambda_{l}).
\label{eq:avFQmath2_physics}
\end{equation}
This can be rearranged as sum of a term containing the von Neumann entropy and another term with the logarithms of the eigenvalues as
\begin{equation}
\AVGFQLOG{A}=
-\frac{2}{N_{\rm g}}\left(2dS+2\sum_k \log \lambda_k\right),
\label{eq:avFQmath2_physics2}
\end{equation}
where the von Neumann entropy $S$ is given in \EQ{eq:defS}.

Next, we ask what the minimal value of $\AVGFQLOG{A}$ is for a given value of $S.$ This can be determined by minimizing \EQ{eq:avFQmath2_physics2} with some constraints. The constraints of $\sum_{k=1}^{d} \lambda_k=1$ can be taken into account by minimizing
\bea\label{eq:fff}
f(\vec\lambda)&=&\frac{2}{N_{\rm g}}\left(2d\sum_{k=1}^d\lambda_{k}\log \lambda_{k}-
2\sum_{k=1}^d\log \lambda_{k}\right),
\eea
where $\vec\lambda=\{ \lambda_k \}_{k=1}^{d-1}$ and we set $\lambda_d:=1-\sum_{k=1}^{d-1} \lambda_k.$ 
Considering the other constraint for the entropy, we arrive at 
\bea\label{eq:ggg}
g(\vec\lambda,\mu_1)&=&\frac{4}{N_{\rm g}}\Bigg[-dS_0-
\sum_{k=1}^d\log \lambda_{k}\nonumber\\
&-&\mu \Bigg(-\sum_{k,l=1}^d\lambda_{k}\log \lambda_{k}-S_0\Bigg)\Bigg],
\eea
where $\mu$ is a Lagrange multiplier and $S_0$ is a constant. 
The allowed region for $\vec \lambda$ is determined by the conditions $\lambda_k\ge 0$ for $k=1,2,...,d-1$ and $\sum_{k=1}^{d-1}\lambda_k\le 1.$

We are looking for the $\vec\lambda$ that minimizes $g$ for some $\mu.$ 
In principle, the minimum could be taken on the boundary of the allowed region for $\vec\lambda.$ However, $\lambda_k\rightarrow +0$ leads to $g\rightarrow\infty,$ hence a minimum cannot be obtained this way.
The other possibility is that the minimum is taken at  the $\vec\lambda$ that fulfills
\begin{subequations}\label{eq:deriv0}
\bea
\frac{\partial g}{\partial \lambda_k}&=&0 \text{ for } k=1,2,..,d-1, \label{eq:deriv0a}\\
\frac{\partial g}{\partial \mu}&=&0.\label{eq:deriv0b}
\eea
\end{subequations} 
From the condition that the derivatives with respect to $\lambda_k$ are zero, \eqref{eq:deriv0a}, follows 
\be\label{eq:lambdalambdaloglog}
\frac1 {\lambda_k}-\frac1 {\lambda_d}-\mu\left(\log \lambda_k-\log \lambda_d\right)=0
\ee
for $k=1,2,..,d-1.$ For a given $k,$  $\lambda_k=\lambda_d$ is clearly a solution.
For some $\mu$ values, there is a second solution.
Then, the $\vec \lambda$ satisfying \EQ{eq:deriv0a} has the following properties.
A possibility is that all elements of $\vec \lambda$ are equal to each other
\bea
\lambda_k&=&(1-\lambda_d)/(d-1) \text{ for } k=1,2,..,d-1
 \label{eq:lambdamin}
\eea
Another possibility is that some of the elements of $\vec \lambda$ are equal to $\lambda_d.$ The other elements are different from $\lambda_d,$ and they are all equal to each other.

\begin{figure}
\centerline{ 
\epsfxsize4cm \epsffile{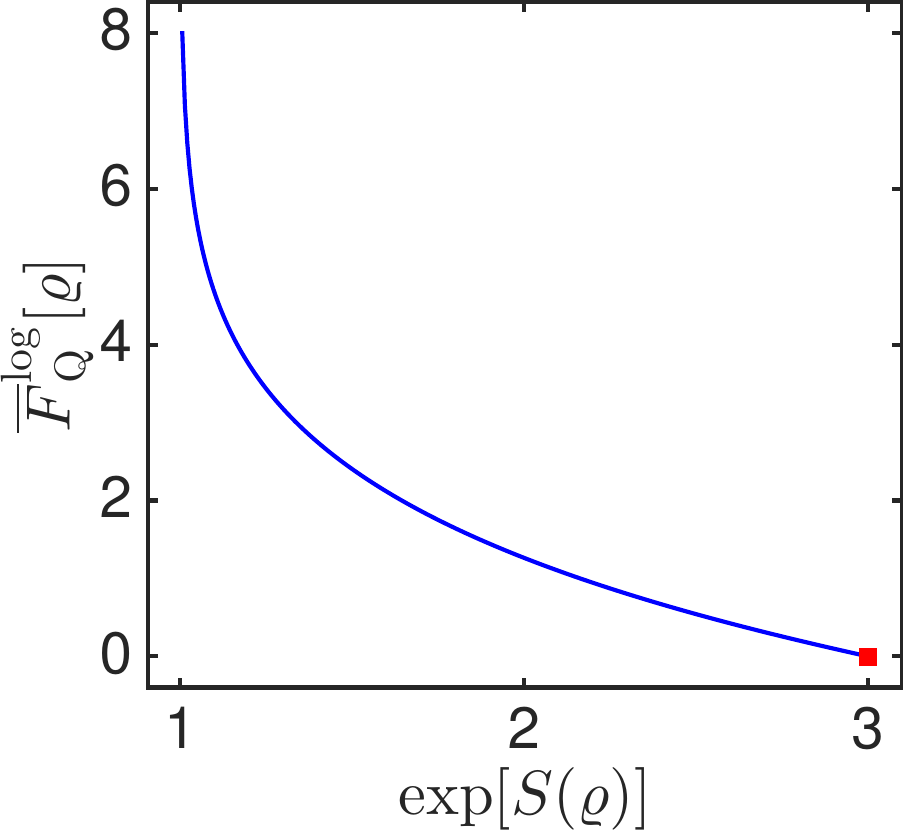}\hskip0.4cm
\epsfxsize4.15cm \epsffile{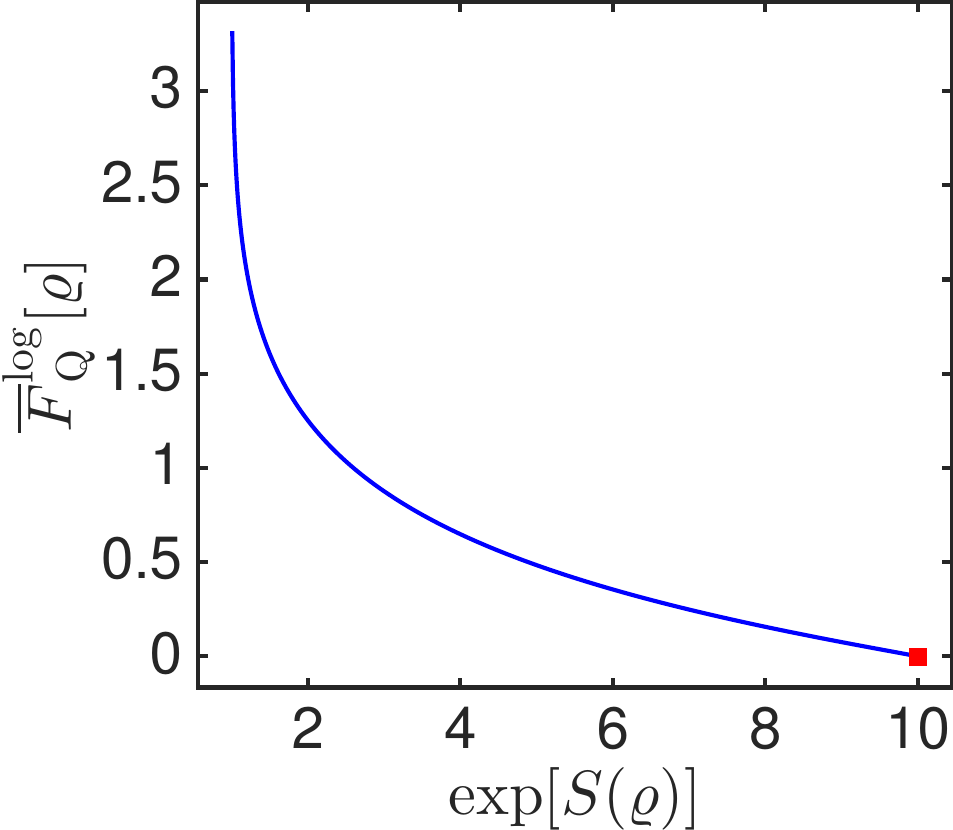}
}
\caption{The relation between the von-Neumann entropy  $S(\varrho)$  and the average $F^{\rm log}[\varrho,A]$ defined in \EQ{eq:avFQmath2_physics2}  for (left) $d=3$ and (right) $10.$  (solid) Points corresponding to states with eigenvalues given in \EQ{eq:lambdamin}. No point corresponding to any  quantum state can be below this line. (square) Completely mixed state \eqref{eq:rhocm}.}
\label{fig:FQlog}
\end{figure}

So far we looked for $\vec \lambda$ for which the derivative of $g$ is zero, which is only a necessary condition for obtaining a minimum of $f(\vec \lambda)$ with the given constraints.
Simple calculations show that the $\vec \lambda$ given in \EQ{eq:lambdamin} minimizes  $f(\vec \lambda)$ for a given value of the von Neumann entropy if $1/d\le \lambda_d \le 1.$  
Such eigenvalues correspond to a quantum state that is a mixture of a pure state and white noise.
The average quantum Fisher information $\AVGFQLOG{A}$ as a function of $\exp(S)$  corresponding to the eigenvalues given in \EQ{eq:lambdamin} is plotted in \FIG{fig:FQlog}.

The results of this section complement the results of \REF{Huber2017Geometric}, where they established a quantum version of the classical isoperimetric inequality relating the quantum Fisher information and the entropy power of a quantum state. They studied multi-mode continuous variable systems, where the averaging was carried out for the canonical operators $x_k$ and $p_k.$ The quantum Fisher information was defined based on a second order derivative of the relative entropy, just as in our last example \EQ{eq:FlogSrel}. In contrast, we considered systems of finite dimension, and averaged the quantum Fisher information over all Hermitian observables. 

\section{Conclusions}

We considered a generalized variance defined as the difference between  the variance and the quantum Fisher information over four. 
We obtained lower bounds on it with the purity of the state. 
We also considered the generalized variance averaged over all Hermitian operators. We found that it is a weighted sum of the linear entropy and another simple term that is the sum of the pairwise harmonic means of the eigenvalues of the density matrix. 
We examined the relation of our quantity to the von Neumann entropy. 
We found also relations between the Kubo-Mori-Bogoliubov quantum Fisher information averaged over all Hermitian operators
and the von Neumann entropy.

\acknowledgments

We thank I. Apellaniz, J. Calsamiglia, J. Ko\l ody\'nski, M. Kleinmann, K. Macieszczak, M. Mosonyi, J. Pitrik, A. Sanpera, T. Schulte-Herbr\"uggen, A. Vershynina, D. Virosztek, G. Vitagliano, and A. Winter for discussions.
We acknowledge the support of the 
EU (ERC Starting Grant 258647/GEDENTQOPT, 
CHIST-ERA QUASAR, COST Action CA15220, QuantERA CEBBEC),
the Spanish Ministry of Economy, Industry and Competitiveness and the European Regional Development Fund FEDER through Grant No. FIS2015-67161-P (MINECO/FEDER, EU), the Basque
Government (Project No. IT986-16), the National Research, Development and Innovation Office NKFIH (Contract No. K124351) 
and the UPV/EHU program UFI 11/55.

\appendix

\section{Calculations for averages over the Hermitian operators}
\label{app:avF}

In this Appendix, we prove Observation 4. Let us evaluate the sum of variances over all generators in \EQ{eq:avgvar}. Based on the well-known identities (e.g., see \REF{Toth2015Evaluating})
\begin{subequations}
\bea
\sum_{m=1}^{N_{\rm g}}\exs{(A^{(m)})^{2}}&=&2\left(d-\frac{1}{d}\right),\label{eq:sumA2}\\
\sum_{m=1}^{N_{\rm g}}\exs{A^{(m)}}^{2}&=&2\left[{\rm Tr}(\varrho^{2})-\frac{1}{d}\right],
\eea
\end{subequations}
we obtain for the average variance \EQ{eq:sumvar}.
Note that here we used the normalization given in \EQ{eq:trAkAl}.

Let us obtain an equation for averages of the elements of $A$ explicitly. 
 Based on \EQ{eq:sumA2}, averaging  the second moment of $A_{\vec n}$ can be rewritten as 
\be\label{eq:avgvar22}
\AVGVECN{A}=\AVGAXX{A}+(d-1)\AVGAXY{A},
\ee
where $A_{\vec n}$ is defined in \EQ{eq:An}. We took into account that the averages for diagonal elements are equal to each other, hence
\be
\AVGAXX{A}=\AVGAKK{A}
\ee
holds for all $k.$ Similarly, the averages for off-diagonal elements are also equal to each other, and we obtain 
\be
\AVGAXY{A}=\AVGAKL{A}
\ee
for all $k\ne l.$


After calculating the average of the variance for traceless Hermitian operators, we calculate an analogous quantity for the quantum Fisher information. The quantum Fisher information for a traceless Hermitian operator can be obtained as
\be
F_{\rm Q}[\varrho,A_{\vec n}]=\vec{n}^T F \vec{n},
\ee
where $F$ is the Fisher matrix defined as \cite{Paris2009QUANTUM}
\be 
F_{mn}=F_{\rm Q}[\varrho,A^{(m)},A^{(n)}]
=2\sum_{k,l}\frac{(\lambda_{k}-\lambda_{l})^{2}}{\lambda_{k}+\lambda_{l}} A_{kl}^{(m)} A_{lk}^{(n)}.\label{eq:FQAB}
\ee
Note that \EQ{eq:FQAB} coincides with \EQ{eq:FQ} if $A^{(m)}=A^{(n)}=A.$

Next, we consider the quantum Fisher information averaged over the Hermitian operators \cite{[{It is also possible to average over operators with a given spectrum, which has been used in a different context for bipartite systems, see }] [{.}]Farace2016Building}. The average has been calculated in \REFS{Li2013Entanglement}. Here, we present an alternative proof for completeness, as well as, for proving \EQS{eq:avgAn12} and \eqref{eq:avgA11}.

Based on ideas similar to the ones used for the average variance given in \EQ{eq:avgvar}, we  obtain 
\be
\AVGFQ{A}=\frac{1}{N_{\rm g}}\sum_{m=1}^{d^2-1} F_{\rm Q}[\varrho,A^{(m)}].
\ee
Let us now evaluate the sum of the quantum Fisher information for all generators in \EQ{eq:FQ}.
 Averaging \EQ{eq:FQ} can be rewritten as
\be\label{eq:avgFQAN}
\AVGFQ{A}=\left(2\sum_{k,l}\frac{(\lambda_{k}-\lambda_{l})^{2}}{\lambda_{k}+
\lambda_{l}}\right) \AVGAXY{A}.
\ee
Using the identity
\be
\sum_{k,l}\frac{(\lambda_{k}-\lambda_{l})^{2}}{\lambda_{k}+\lambda_{l}}=
\sum_{k,l}\left(\frac{(\lambda_{k}+\lambda_{l})^{2}}{\lambda_{k}+\lambda_{l}}-\frac{4\lambda_{k}\lambda_{l}}{\lambda_{k}+\lambda_{l}}\right),
\ee
\EQ{eq:avgFQAN} can be further rewritten with $H(\varrho)$ as
\be\label{eq:avgFQ22}
\AVGFQ{A}=4\left[d-H(\varrho)\right] \AVGAXY{A}.
\ee

Next, we  determine $\AVGAXY{A}.$ For that we write down the average quantum Fisher information for pure states in two different ways. On the one hand, from \EQ{eq:avgFQ22} we obtain for a pure state $\ket{\Psi}$ the average quantum Fisher information as
\be\label{eq:avgFQ22b}
\AVGFQPURE{A}=4\left(d-1\right) \AVGAXY{A}.
\ee
On the other hand, we know that for pure states the quantum Fisher information is four times the variance. Hence, using the formula \eqref{eq:sumvar} for the the average variance, for pure states $\ket{\Psi}$ we arrive at
\be\label{eq:avgFQ22c}
\AVGFQPURE{A}=4 \AVGVAR{A}=\frac{8}{N_{\rm g}}(d-1).
\ee
Comparing \EQ{eq:avgFQ22b} and \EQ{eq:avgFQ22c} we arrive at \EQ{eq:avgAn12}
and obtain the average quantum Fisher information as \EQ{eq:sumFQ}.

The quantum Fisher information $F_{\rm Q}[\varrho,A^{(k)}]$ is convex in the state. Hence, the quantum Fisher information averaged over the Hermitian operators \eqref{eq:sumFQ} is also convex in the state. From \EQ{eq:sumFQ} it also follows that  $H(\varrho)$ is concave in the state. \EQL{eq:sumFQ} is maximal for pure states. Hence, the average quantum Fisher information, i.e., the metrological usefulness of the quantum state is the largest for pure states. 

Finally, we prove the formula for the average for the diagonal elements of $A$ given in 
\EQ{eq:avgA11}.
Based on \EQ{eq:var} we know that 
for the completely mixed state, \eqref{eq:rhocm}, for any $A$
\be
\exs{A^2}_{\rho_{\rm cm}}=\frac{1}{d}\sum_{k,l} \vert A_{kl} \vert^2 = 
\frac{1}{d}{\rm Tr}(A^2) \label{eq:A2}
\ee
holds.
Due to the normalization of the basis matrices given in \EQ{eq:trAkAl} we arrive at
\be
\exs{A_{\vec n}^2}_{\rho_{\rm cm}}=\frac{2}{d}\label{eq:A22}
\ee
for any $\vec n.$ 
From \EQ{eq:avgAn12}, \EQ{eq:A22} and \EQ{eq:avgvar22} we obtain \EQ{eq:avgA11}.

\section{The relation between the Kubo-Mori-Bogoliubov Fisher information and the relative entropy}
\label{app:Srel}

In \REF{Petz2010From}
, the generalized quantum Fisher information is defined as
\begin{eqnarray}
F_{\rm Q}^{\log}(\varrho;A)
&=&\sum_{ij}\frac{1}{m_{f}(\lambda_{i},\lambda_{j})}\vert A_{ij} \vert^2,\label{eq:genFQ}
\end{eqnarray}
where $f:\mathbbm{R}^+\rightarrow\mathbbm{R}^+$ is a matrix monotone function,
and a corresponding mean $m_f$ is defined as in \EQ{eq:mean}.
The quantum Fisher information family, \eqref{eq:genFQ}, includes the usual quantum Fisher information for $f(x)=(1+x)/2.$ It also includes the Kubo-Mori-Bogoliubov Fisher information with 
\be 
f(x)=(x-1)/\log(x)
\ee 
defined as
\be
F_{\rm Q}^{\log}(\varrho;A)= \sum_{k,l}
\frac{\log(\lambda_{k})-\log(\lambda_{l})}{\lambda_{k}-\lambda_{l}}
\vert A_{kl} \vert^{2}.\label{eq:FQKBM_lin}
\ee
The quantum Fisher information, \eqref{eq:FQKBM_lin}, corresponds to linear dynamics of the type
\be
\varrho_t=\varrho_0+Bt,
\ee
where $t$ is the parameter of the quantum state and $B$ is a traceless Hermitian matrix.
However, in physics we typically consider a unitary dynamics of the type
\be
\varrho_\theta = e^{-iA \theta}\varrho e^{+iA \theta}.
\ee
The Kubo-Mori-Bogoliubov Fisher information corresponding to such a dynamics can be obtained as
\be
F_{\rm Q}^{\log}[\varrho,A]=F_{\rm Q}^{\log}(\varrho;i[\varrho,A]),
\ee
which is identical to the formula given in \EQ{eq:FQlog_physics}.

\EQL{eq:FlogSrel} can be proved as follows.
In Ref.~\cite{Hayashi2002Two}, it is shown that for small $\theta$
\begin{equation}
F^{\log}[\varrho,A] \approx \frac{2}{\theta^2} S(\varrho\vert\vert \varrho_\theta)
\label{eq:Hayashi}
\end{equation}
holds.
Now, note the trivial relation
\begin{equation}
S(\varrho\vert\vert \varrho_\theta)=0
\end{equation}
for $\theta=0.$
Note also that
\begin{equation}
\frac{d}{d t}S(\varrho\vert\vert \varrho_\theta)\vert_{\theta=0}
=0,
\end{equation}
since $S(\varrho\vert\vert \varrho_\theta)$ is minimal for $\theta=0.$ Hence,
\begin{equation}
S(\varrho\vert\vert \varrho_\theta)\approx \frac{\theta^2}{2}\frac{d^2}{d^2 \theta}S(\varrho\vert\vert \varrho_\theta) + O(\theta^3).\label{eq:Staylor}
\end{equation}
From Eqs.~\eqref{eq:Hayashi} and \eqref{eq:Staylor} follows Eq.~\eqref{eq:FlogSrel}.
On the relation between the Kubo-Mori-Bogoliubov Fisher information and the relative entropy, see also
\REFS{Petz2002Covariance,Konig2014The}.

\bibliography{Bibliography2.bib}

\end{document}